# The Spectral Domain Snell's Law in Diffusion-Wave Fields

Pengfei Zhu,[1,*] Julien Lecompagnon,[1] Philipp Daniel Hirsch,[1] Mathias Ziegler,[1] Andreas Mandelis[2,3,†]

[1]*Bundesanstalt für Materialforschung and -prüfung (BAM)*, 12205 Berlin, Germany
[2]*Center for Advanced Diffusion-Wave and Photoacoustic Technologies (CADIPT), Department of Mechanical and Industrial Engineering, University of Toronto*, Toronto M5S 3G8, Canada
[3]*Institute for Advanced Non-Destructive and Non-Invasive Diagnostic Technologies (IANDIT), Faculty of Applied Science and Engineering, University of Toronto*, Toronto M5S 1A4, Canada

Snell's law is traditionally regarded as a hallmark of phase-propagating phenomena such as optical, acoustic, elastic, electromagnetic, and quantum waves. In contrast, purely diffusive processes – such as Fourier heat conduction and chemical diffusion – are generally considered incapable of exhibiting refractive/reflective behavior. In this letter, we demonstrate that although diffusion waves including thermal diffusion, mass diffusion, Lindblad quantum diffusion, and electromagnetic diffusion do not follow Snell's law in either time- or frequency-domain, nevertheless they obey a spectral form of Snell's law which reveals a hidden analog of wave refraction/reflection within the mathematical structure of diffusion dynamics. Remarkably, the spectral refraction ratio is governed not by the diffusion coefficient itself but by the constitutive relations of the media across the interface, establishing a new physical paradigm for diffusion-wave fields. Importantly, while each spectral eigenmode satisfies a rigorous Snell-type refraction relation, the inverse Fourier-Laplace transformation mixes these modes and suppresses any persistent real-space refraction angle, thereby reconciling the modal-level directionality with the long-standing absence of geometric refraction in diffusive systems.

*Introduction*—In recent years, several diffusion-related periodic phenomena have become unified under the global mathematical label of diffusion-wave field [1,2], mainly including the categories of charge-carrier-density waves [3], diffuse photon density waves [4], mass-transport waves [5], thermal waves [6], chemical waves [7], etc. Unlike hyperbolic traveling waves such as acoustic or optical waves, the physics of diffusive thermal waves is governed by the parabolic heat diffusion equation and in accordance to Fickian principles [8]. The description of thermal diffusion-wave propagation depends on the spatial and temporal scale. Classic Fourier's law predicts the thermal wave has single characteristic curve with infinite propagation speed thereby cannot explain ultrafast non-Fourier heat conduction under low-dimensional structures, or ballistic heat transfer. The Cattaneo-Vernotte model [9,10] introduces finite propagation time constant $\tau$ (thermal inertia), which modifies the "infinite speed" non-physical phenomena and can be used to describe thermal wave or second sound phenomena. However, it still lacks the ability to capture ballistic transport [11,12]. Other descriptions such as the Boltzmann transport equation (BTE) [13] and its modified forms [14,15] significantly depend on empirical parameters, steady-state assumption, and complex and costly computation. The argument of a wave-like behavior in diffusive phenomena arises from the similarity between frequency-domain thermal diffusion equation (hyperbolic Helmholtz equation) and wave equation. The only difference is the wave number $k$ becomes complex. Many experiments demonstrate the periodic temperature fluctuations that attenuate with depth and exhibit phase delays in modulated thermography [16,17]. Therefore, Bertolotti et al. [18] tried to use optics concepts (refraction and reflection) to describe these experimental observations. They selected a single frequency modulated heat source to solve the heat conduction equation and obtained the time-domain "Snell's law" in thermal diffusion-wave fields. Furthermore, O'Leary et al [4] experimentally validated the frequency-domain Snell's law in diffuse photon density waves. However, Mandelis et al. [19] strictly criticized this kind of time-domain wave theory as applied to parabolic thermal diffusion-wave fields exhibiting a single characteristic curve $\xi(t) = t = $ const with propagation velocity $v = \frac{dx}{dt} = \infty$. Furthermore, they mathematically derived the "reflection and refraction" behavior of thermal diffusion-wave fields at the interface under a harmonic heat source. The results yielded a complex function related to the radial position $r$ rather than a constant. Mandelis et al. attributed Snell's law in experimental results of diffuse photon density waves as an abnormal phenomenon under specific incident angle. Thus, it was shown that Snell's law in thermal diffusion-wave fields does not hold in either time domain or frequency domain.

To further reveal the physical aspects of interfacial relationships in diffusion-wave fields, we consider the case of thermal diffusion waves with all possible two-dimensional sources in a thermally isotropic half-space $0 \leq z < \infty$ of thermal conductivity $K_1$ undergoing boundary interactions along the plane $z = 0$, as shown in Fig. 1. The boundary value problem (BVP) can be formulated as:

$$\partial_t u_1(x,z,t) = \alpha_1(\partial_{xx}u_1 + \partial_{zz}u_1) + f_1(x,z,t), \quad z > 0 \quad (1)$$

$$\partial_t u_2(x,z,t) = \alpha_2(\partial_{xx}u_2 + \partial_{zz}u_2) + f_2(x,z,t), \quad z < 0 \quad (2)$$

where $u_j(x,z,t)$ denotes the temperature field in each layer ($j = 1,2$), $\alpha_j$ is the thermal diffusivity, and $f_j(x,z,t) = \delta(x - x_0)\delta(z - z_0)\delta(t)$ is the impulsive heat source. At interface $z = 0$, temperature and heat flux are continuous, thus the boundary conditions are:

$$u_1(x,0,t) = u_2(x,0,t) \quad (3)$$
$$K_1 \partial_z u_1(x,0,t) = K_2 \partial_z u_2(x,0,t) \quad (4)$$

where $K_j$ is the thermal conductivity of layer $j$. The tangential invariance in Snell's law, combined with the wave-like form of the heat equation in the frequency domain (i.e., its resemblance to a Helmholtz equation), motivates the use of a spatial Fourier transform ($x \to k$) and a temporal Laplace transform ($t \to s$):

$$\tilde{u}_j(k,z,s) = \int_0^\infty e^{-st}\left(\int_{-\infty}^\infty e^{ikx} u_j(x,z,t)dx\right)dt \quad (5)$$

where the Laplace variable $s$ acts as a generalized frequency, and $k$ is the spatial frequency. The transformed equation becomes:

$$\alpha_j\left(\partial_{zz}\tilde{u}_j - \sigma_j^2 \tilde{u}_j\right) = -\tilde{f}_j(k,z,s) \quad (6)$$

where $\sigma_j$ is a complex thermal wave number, $\sigma_j^2(k,s) = k^2 + \frac{s}{\alpha_j}$, encoding both spatial decay and phase delay of thermal diffusion. The principal branch is chosen such that $\Re(\sigma_j) > 0$ to ensure decaying behavior as $z \to \pm\infty$. This guarantees the physical boundedness of thermal diffusion fields in both layers. The diffusion-wave field distribution in real space and spectral space is shown in Fig. 1. As mentioned before, the source is located at layer 1 ($0 \leq z < \infty$), thus $\tilde{f}_1 = e^{-ikx_0}\delta(z - z_0)$ and $\tilde{f}_2 = 0$. In source-free regions of each layer, the solution satisfies the homogeneous ordinary differential equation (ODE):

$$\partial_{zz}\tilde{u}_j - \sigma_j^2 \tilde{u}_j = 0 \quad (7)$$

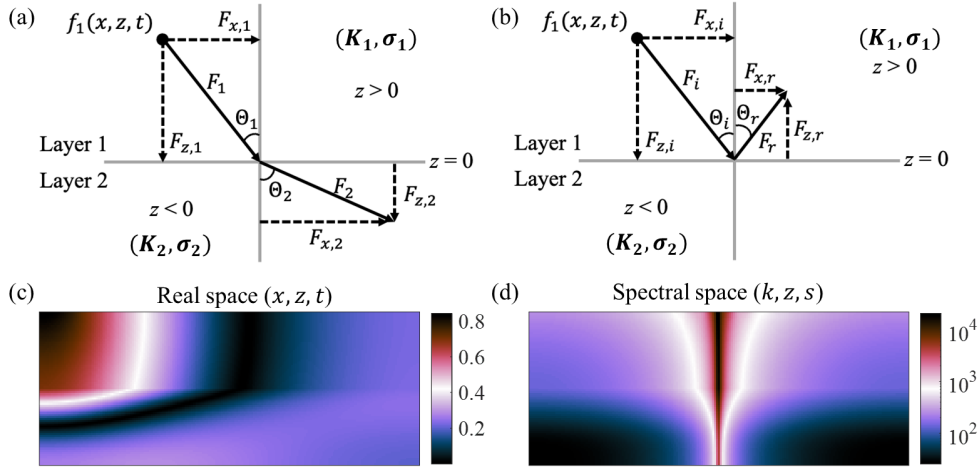

FIG. 1. Interface diffusion-wave flux vector relations under accumulation conditions: (a) Refraction; (b) Reflection; (c) Diffusion-wave field distribution in real space; (d) Diffusion-wave field distribution in spectral space.

The above equation is a constant-coefficient second-order ODE, and the solution is a homogeneous linear combination $A_j e^{\sigma_j z} + B_j e^{-\sigma_j z}$. The integration constants must satisfy the physical boundedness constraint: 1) For $z \to +\infty$, only the decaying term $e^{-\sigma_1 z}$ is physical; 2) For $z \to -\infty$, only $e^{\sigma_2 z}$ remains finite. Furthermore, at the source $z = z_0$ in layer 1, the jump conditions induced by delta-source should be satisfied, the latter representing an instantaneous point heat injection source at $(x_0, z_0)$. Integrating Eq. (6) across $z = z_0$ gives the jump condition:

$$\partial_z \tilde{u}_j|_{z=z_0^+} - \partial_z \tilde{u}_j|_{z=z_0^-} = -\frac{1}{\alpha_1} e^{-ikx_0} \quad (8)$$

We also have continuous conditions across the interface:

$$\tilde{u}_1(k,0,s) = \tilde{u}_2(k,0,s) \quad (9)$$
$$K_1 \partial_z \tilde{u}_1(k,0,s) = K_2 \partial_z \tilde{u}_2(k,0,s) \quad (10)$$

Solving the interface conditions Eq. (9) and (10) yields:

$$R(k,s) = \frac{K_1\sigma_1 - K_2\sigma_2}{K_1\sigma_1 + K_2\sigma_2} \quad (11)$$
$$T(k,s) = \frac{2K_1\sigma_1}{K_1\sigma_1 + K_2\sigma_2} \quad (12)$$

where $R(k,s)$ and $T(k,s)$ are the spectral reflection and transmission coefficients for diffusive waves. Their magnitude and phase describe how each spectral component is reflected or refracted. The analytical solution for $z > 0$ (layer 1) is:

$$\tilde{u}_1(k,z,s) = \frac{e^{-ikx_0}}{2\alpha_1\sigma_1}[e^{-\sigma_1|z-z_0|} + R(k,s)e^{-\sigma_1(z+z_0)}] \quad (13)$$

where the first term in Eq. (13) is the incident field and the second terms represents is the coherently accumulated or depleted interfacial field, usually identified by the term "reflected field" normally assigned to propagating wave-fields. The analytical solution for $z < 0$ (layer 2) is:

$$\tilde{u}_2(k,z,s) = \frac{e^{-ikx_0}}{2\alpha_1\sigma_1}T(k,s)e^{-\sigma_1 z_0}e^{\sigma_2 z} \quad (14)$$

Eq. (14) represents the refracted component into layer 2. The physical-space temperature is recovered via inverse Fourier ($k \to x$) and Laplace transforms ($s \to t$):

$$u_j(x,z,t) = \frac{1}{2\pi i}\int_\Gamma e^{st}[\frac{1}{2\pi}\int_{-\infty}^{\infty}\tilde{u}_j(k,z,s)e^{ikx}dk]ds \quad (15)$$

where $\Gamma$ is the Bromwich path of inverse Laplace transform. The heat flux is defined by Fourier's law:

$$\mathbf{F}_j(x,z,t) = -K_j\nabla u_j(x,z,t) = (F_{x,j}, F_{z,j}) \quad (16)$$

where $F_{x,j} = -K_j\partial_x u_j$, $F_{z,j} = -K_j\partial_z u_j$. $\mathbf{F}_j$ describes both magnitude and direction of energy flow and its direction defines a "diffusion angle" relative to the normal, as shown in Fig. 1(a). The definition of spectral "incident angle" is:

$$\tan\tilde{\Theta}_j^{(sp)}(k,s) = \frac{\tilde{F}_{x,j}(k,0^\pm,s)}{\tilde{F}_{z,j}(k,0^\pm,s)} \quad (17)$$

where we label upper $u_1$ $0^+$, and lower $u_2$ $0^-$. Equation (17) defines the local inclination of heat flux relative to interface normal in the spectral domain, analogous to the angle of incidence or refraction in optics. For the upper layer, we have:

$$\tilde{F}_{x,1}(k,s) = -K_1(ik)\tilde{u}_1(k,0^+,s)$$
$$= -K_1(ik)\frac{e^{-ikx_0}}{2\alpha_1\kappa_1}e^{-\sigma_1 z_0}(1+R(k,s)) \quad (18)$$
$$\tilde{F}_{z,1}(k,s) = -K_1\partial_z\tilde{u}_1(k,z,s)|_{z=0^+}$$
$$= -K_1\frac{e^{-ikx_0}}{2\alpha_1}e^{-\sigma_1 z_0}(1-R(k,s)) \quad (19)$$

The factors $1 \pm R$ arise from the superposition of incident and reflected waves. The "+" corresponds to even (lateral) symmetry, and the "-" captures normal component interference. The spectral flux ratio is:

$$\tan\tilde{\Theta}_1^{(sp)}(k,s) = \frac{\tilde{F}_{x,1}(k,s)}{\tilde{F}_{z,1}(k,s)} = \frac{ik}{\sigma_1}\cdot\frac{K_1\sigma_1}{K_2\sigma_2} = i\frac{kK_1}{K_2\sigma_2(k,s)} \quad (20)$$

$$\tan\tilde{\Theta}_2^{(sp)}(k,s) = \frac{\tilde{F}_{x,2}(k,s)}{\tilde{F}_{z,2}(k,s)} = i\frac{k}{\sigma_2(k,s)} \quad (21)$$

where $\frac{1+R}{1-R} = \frac{K_1\sigma_1}{K_2\sigma_2}$. These angles are generally complex because diffusion is dissipative. Their real parts correspond to the phase inclination (the direction of the phase gradient), while their imaginary parts represent exponential attenuation along that direction. Consequently, they cannot be interpreted as geometric angles in the conventional sense, as in lossless optical systems. Physically, if one wishes to define an effective inclination, it should be taken as $\Re(\tan\tilde{\Theta})$, or, equivalently, derived from the real part of the phase gradient, which indicates the actual direction of energy flow. In this framework, the angles describe the spectral ratio of tangential to normal components of the energy flux, rather than a geometric propagation path. The corresponding spectral Snell's law for diffusion waves can then be expressed as:

$$\frac{\tan\tilde{\Theta}_1^{(sp)}(k,s)}{\tan\tilde{\Theta}_2^{(sp)}(k,s)} = \frac{(\tilde{F}_{x,1}/\tilde{F}_{z,1})}{(\tilde{F}_{x,2}/\tilde{F}_{z,2})} = \frac{K_1}{K_2} \quad (22)$$

Eq. (22) is the refractive law for thermal diffusion waves – formally analogous to Snell's law, but in the spectral (complex) domain. It states that the ratio of spectral heat-flux angles is governed by the ratio of thermal conductivities. The appearance of thermal conductivity rather than diffusivity reflects that the refraction law is formulated in terms of energy flux continuity, not field continuity. Now, let's consider the reflective phenomenon, as shown in Fig. 1(b). The reflected part of the thermal diffusion-wave field in Eq. (13) is:

$$\tilde{u}_r(k,z,s) = \frac{e^{-ikx_0}}{2\alpha_1\sigma_1}R(k,s)e^{-\sigma_1(z+z_0)} \quad (23)$$

According to the definition of heat flux in Fourier's law ($\mathbf{F}_j(x,z,t) = -K_j\nabla u_j(x,z,t)$), we can calculate the reflected flux components in $x$ and $z$ directions:

$$\tilde{F}_{x,r}(k,s) = -K_1(ik)\tilde{u}_r = -K_1(ik)\frac{e^{-ikx_0}}{2\alpha_1\sigma_1}R(k,s) \quad (24)$$

$$\tilde{F}_{z,r}(k,s) = -K_1\partial_z\tilde{u}_r = K_1\sigma_1\frac{e^{-ikx_0}}{2\alpha_1\sigma_1}R(k,s) \quad (25)$$

Therefore, the spectral reflection angle is:

$$\tan\tilde{\Theta}_r^{(sp)}(k,s) = \frac{\tilde{F}_{x,r}}{\tilde{F}_{z,r}} = -i\frac{k}{\sigma_1(k,s)} \quad (26)$$

This has the opposite sign of the incident flux component in $z$, meaning that the normal component reverses direction – consistent with mirror reflection.

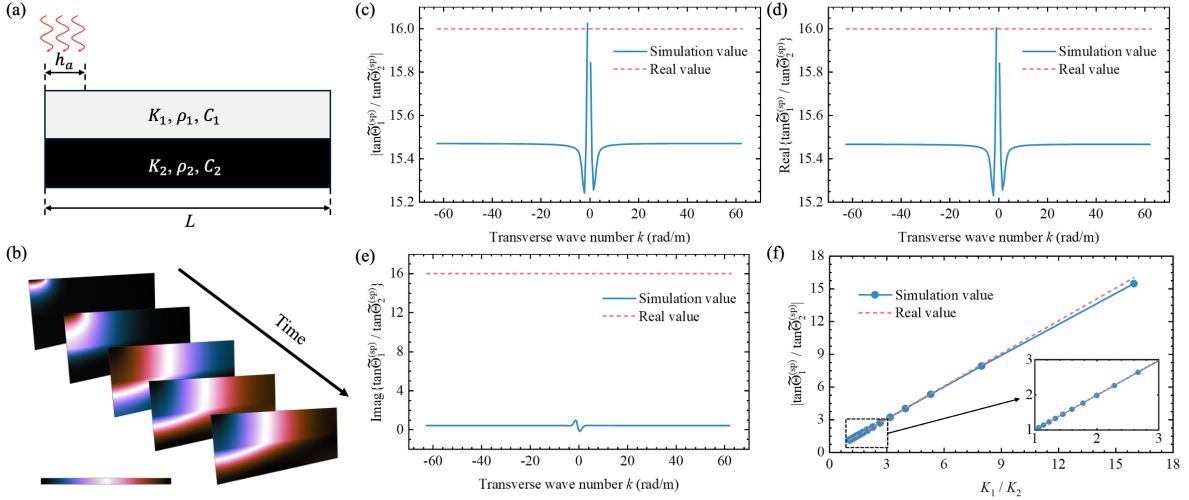

FIG. 2. Simulation validation of spectral domain Snell's law under refractive condition: (a) The model of simulation validation for spectral domain Snell's law under refractive condition; (b) The temperature distribution varying with time; (c) The absolute value of $\tan\widetilde{\Theta}_1^{(sp)}(k,s)/\tan\widetilde{\Theta}_2^{(sp)}(k,s)$; (d) The real part of $\tan\widetilde{\Theta}_1^{(sp)}(k,s)/\tan\widetilde{\Theta}_2^{(sp)}(k,s)$; (e) The imaginary part of $\tan\widetilde{\Theta}_1^{(sp)}(k,s)/\tan\widetilde{\Theta}_2^{(sp)}(k,s)$; (f) The mean absolute values of $\tan\widetilde{\Theta}_1^{(sp)}(k,s)/\tan\widetilde{\Theta}_2^{(sp)}(k,s)$ for different mediums.

The incident spectral component is:

$$\tilde{u}_i \propto \frac{e^{-ikx_0}}{2\alpha_1\sigma_1} e^{-\sigma_1|z-z_0|} \quad (27)$$

Hence,

$$\tan\widetilde{\Theta}_i^{(sp)}(k,s) = +i\frac{k}{\sigma_1(k,s)} \quad (28)$$

This defines the inclination of the incident spectral component relative to the interface. The reflection symmetry is:

$$\tan\widetilde{\Theta}_r^{(sp)}(k,s) = -\tan\widetilde{\Theta}_i^{(sp)}(k,s) \quad (29)$$

Eq. (29) exhibits that each spectral component $(k,s)$ undergoes mirror-like reflection at the interface. The normal component of the flux reverses its sign, while the tangential component retains the same magnitude but opposite direction – consistent with the classical reflection law ("angle of reflection equals angle of incidence") in the spectral domain. The reflection coefficient $R(k,s)$ governs both amplitude and phase: $|R|$ represents the magnitude of reflected energy, whereas $\arg(R)$ denotes the phase delay upon reflection.

Several limiting cases can be identified: 1) When $K_2 \to 0$ (i.e., the layer 2 behaves as a nearly adiabatic medium), $R \to 1$, corresponding to total reflection. In this limit, heat accumulates near the interface, resulting in a slow temporal decay; 2) When $K_1 = K_2$ and $\alpha_1 = \alpha_2$ (identical media), $R \to 0$, implying perfect continuity of the field and the absence of reflection. Mathematically, a "spectral Snell's law" can be defined in the frequency domain because, for each fixed frequency, $\sigma_j$ acts as a (complex) propagation constant that allows boundary matching across the interface. However, upon inverse transformation, the frequency-dependent phase and attenuation terms become coupled. This coupling, together with the strong suppression of high-frequency, highly directional components, produces a time-domain response that is smooth, diffusive, and effectively isotropic in the long-time limit. Consequently, no Snell-type geometric refraction can survive in the time domain.

*Numerical Analysis*—To validate the accuracy of the proposed spectral domain Snell's law in Eq. (22), the finite element model was employed, as shown in Fig. 2(a). The local heating area is $h_a = 4$ mm, the length of the sample is $L = 40$ mm, the thermal conductivity, density, and heat capacity of the upper medium is $K_1 = 80$ W m$^{-1}$K$^{-1}$, $\rho_1 = 4500$ kg m$^{-3}$, and $C_1 = 477$ J kg$^{-1}$K$^{-1}$, the thermal conductivity, density, and heat capacity of the downer medium is, respectively, $K_2 = 75$ W m$^{-1}$K$^{-1}$, $\rho_2 = 8000$ kg m$^{-3}$, and $C_2 = 500$ J kg$^{-1}$K$^{-1}$. The boundary condition in the heated area is $\frac{\partial T}{\partial n}|_{\Gamma = h_a} = K_1 q(t)$, where $q(t) = 1 \times 10^5 \times \sin(2\pi f t)$ W m$^{-2}$, and $f$ is 0.5 Hz. For all other boundaries, we apply a generalized Newton-type heat loss condition $\frac{\partial T}{\partial n}|_{\Gamma \neq h_a} = h(T - T_\infty)$, where $T_\infty = 20$ °C is the ambient temperature. Importantly, the coefficient $h = 10$ W m$^{-2}$K$^{-1}$ represents an effective heat-loss coefficient that accounts for both/either weak convection and linearized radiative losses. In quiescent media, the dominant mechanism is radiative exchange described by the Stefan-Boltzmann law $\sigma T^4$; linearization around $T_\infty$ yields a Newton-type boundary condition identical in form to the one used here. Thus the chosen $h$ captures the combined effects of convection and radiation in the temperature range relevant to our simulations. The simulation results of temperature distribution varying with time are shown in Fig. 2(b). According to Eqs. (18) and (19), it is possible to calculate the heat flux vectors in the spectral domain, $\tilde{F}_{x,1}(k,s)$, $\tilde{F}_{z,1}(k,s)$, and $\tilde{F}_{x,2}(k,s)$, $\tilde{F}_{z,2}(k,s)$. Then, the spectral flux ratios $\tan\tilde{\Theta}_1^{(sp)}$ and $\tan\tilde{\Theta}_2^{(sp)}$ can be calculated based on Eq. (20) and (21). Finally, it is possible to compare the calculated spectral flux ratio $\tan\tilde{\Theta}_1^{(sp)}/\tan\tilde{\Theta}_2^{(sp)}$ and the pre-known $K_1/K_2$. The simulation results are shown in Fig. 2, parts (c) – (f). We compared different calculation results for spectral flux ratios by taking absolute value $|\tan\tilde{\Theta}_1^{(sp)}/\tan\tilde{\Theta}_2^{(sp)}|$ (see Fig. 2(c)), real part $\text{Re}\{\tan\tilde{\Theta}_1^{(sp)}/\tan\tilde{\Theta}_2^{(sp)}\}$ (see Fig. 2(d)), and imaginary part $\text{Im}\{\tan\tilde{\Theta}_1^{(sp)}/\tan\tilde{\Theta}_2^{(sp)}\}$ (see Fig. 2(e)).

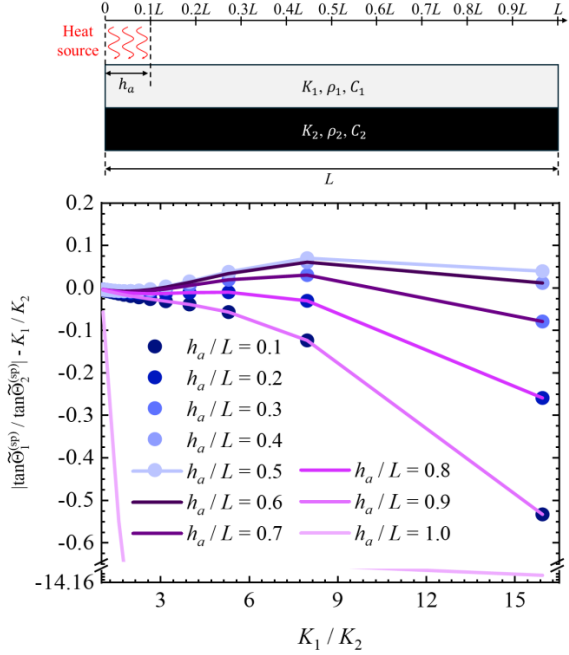

FIG. 3. Simulation validation for spectral domain Snell's law under different heat source widths. The curve corresponding to $h/L = 1$ is truncated to highlight its deviation from the theoretical prediction at large $K_1/K_2$.

According to the simulation results, the absolute value and real part present adherence to the proposed spectral Snell's law in thermal diffusion-wave fields. It is noted that the spectral Snell's law is stricter than that in wave matter as it requires that each frequency component should be the same, equal to the thermal conductivity ratio $K_1/K_2$ rather than a single component. The absolute value of the spectral flux ratio at each frequency component is close to the real value, as shown in Figs. 2(c), (d), and (e). Under localized heating, the thermal excitation has a broad spectrum of lateral spatial frequencies. Each $k_x$ component excites a distinct thermal diffusion mode with different attenuation and transmission characteristics across the interface. As a result, the spectral ratio is inherently a $k_x$-dependent function rather than a constant (see Figs. 2(c), (d), and (e)). Moreover, due to symmetry and gradient selection rules, all directional spectral quantities vanish at $k_x = 0$, leading to a zero response at normal incidence and a peaked structure at finite spatial frequencies. The above results are focused on a fixed thermal conductivity ratio. To validate the accuracy of the proposed spectral domain Snell's law, the thermal conductivity of the bottom medium was changed from 5 to 75 with an interval of 5. The simulation results are shown in Fig. 2(f). It can be seen that the simulation results are in good agreement with the real part of the complex ratio in Eq. (22) which is equal to the ratio of thermal conductivities $K_1/K_2$, especially in the low ratio value range. Therefore, it is concluded that in the spectral domain, Snell's law has been validated under local heating.

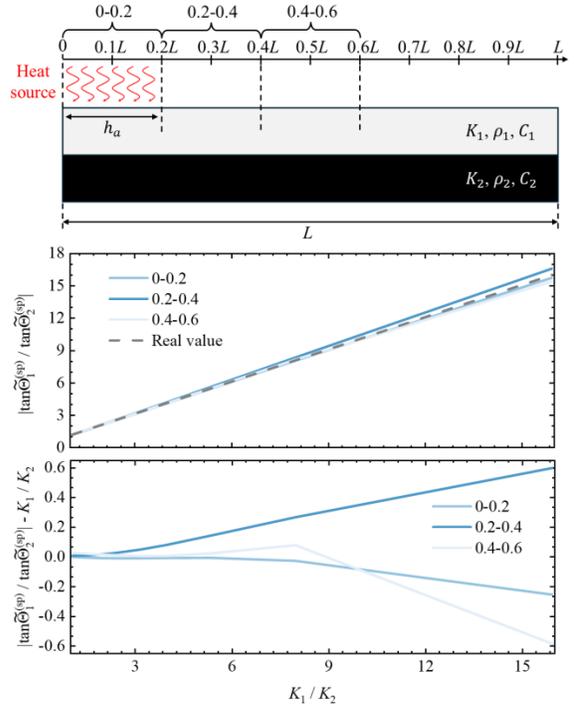

FIG. 4. Simulation validation for spectral domain Snell's law under different heat source locations.

In the previous simulation, we considered a local heat source with finite width. Now, the influence of heat source width, i.e., the heating area $h_a$ will be discussed. The ratio of heating area and sample length $h_a/L$ was set from 0.1 to 1.0 (uniform full-field heating). The simulation results are shown in Fig. 3. Except for 1.0 (uniform full-field heating), all simulation results have good agreement with the real part of Eq. (22), i.e. the ratio of thermal conductivities $K_1/K_2$, thereby once again validating the accuracy of the proposed spectral domain Snell's law. The reason why the spectral domain Snell's law fails under uniform full-field heating is that the spectral domain Snell's law relies on the conservation of lateral wavevector components, which are selected by the spatially localized heating. Under uniform full-field heating, the excitation contains only the zero lateral spatial frequency ($k_x = 0$), and the thermal diffusion process degenerates into a purely one-dimensional normal diffusion. Consequently, the notion of refraction and the associated spectral domain Snell's law lose their physical meaning. Furthermore, it is clear that the spectral flux ratio $\tan\tilde{\Theta}_1^{(sp)} / \tan\tilde{\Theta}_2^{(sp)}$ is symmetrical around half of the ratio of heating area and sample length ($h_a/L = 0.5$). Therefore, the experimental results show that the $h_a/L = 0.4$ is equal to $h_a/L = 0.6$, $h_a/L = 0.3$ is equal to $h_a/L = 0.7$, $h_a/L = 0.2$ is equal to $h_a/L = 0.8$, and $h_a/L = 0.1$ is equal to $h_a/L = 0.9$. The simulation results validate that the proposed spectral domain Snell's law is not affected by the heat source width.

Finally, it is necessary to validate the influence of the heat source location. The sample length was divided into equal segments at $h_a/L = 0.2$ intervals, and the width of the heat source was fixed and set to $h_a/L = 0.2$. Therefore, due to the symmetry, we only need to consider three heat source locations, i.e., $h_a/L = 0\sim0.2$, $h_a/L = 0.2\sim0.4$, and $h_a/L = 0.4\sim0.6$. The corresponding simulation results are shown in Fig. 4. We find that the retrieved refraction angles remain in good agreement with the theoretical values throughout the domain, confirming that the proposed spectral-domain Snell's law is robust against variations in source position. It is worth noting that earlier diffusion-wave studies both in photon [4] and thermal [19] fields have reported increased anomalous refraction when the thermal-wave source approaches the interface, a behavior attributed to interface-interactive coherent accumulation or depletion of thermal energy induced by diffusive losses in the frequency-domain propagation dynamics. In the present work, however, the formulation was fully carried out in the spectral domain, where such diffusive effects enter implicitly through the complex wavenumber $\sigma(k,s)$. As a result, the enhanced anomaly near the interface is substantially attenuated once transformed into the spectral domain, where the steady-state harmonic field automatically incorporates diffusive losses through the imaginary component of $\sigma(k,s)$. Within this spectral-domain framework, the influence of source proximity to the interface remains small in the parameter range considered, and the effective refraction rule retains its predictive accuracy.

The above observation provides an important physical clarification regarding the coexistence of Snell's law invalidity in the time/frequency domains and its formal validity in the spectral domain. In real space or in single-frequency representations, the thermal field results from the superposition of a continuum of $(k,s)$ components, each possessing different attenuation rates. This spectral mixing destroys any persistent geometric directionality, preventing the emergence of a unique refraction angle – hence the breakdown of Snell's law in both the time domain and the temporal frequency domain. In contrast, in the spectral domain the diffusion operator becomes diagonal, and each $(k,s)$ component behaves as an independent eigenmode with a well-defined flux direction. For each such mode, continuity of temperature and normal heat flux enforces a rigorous refraction condition, yielding the spectral-domain Snell's law derived in Eq. (22). Thus, the mathematical origin of Snell-like behavior lies entirely at the modal level; its disappearance in real space is a direct consequence of dissipative modal superposition. This insight has several consequences for the treatment of diffusion-wave transport at interfaces. First, it provides a unified directional framework for predicting interfacial flux partitioning in layered diffusive media – something unattainable in the time or frequency domains alone. Second, although a geometric refraction diagram cannot be reconstructed after inverse transformation (because modal directions do not survive the superposition), the spectral Snell relation offers a practical route to constructing modal refraction maps that can guide the interpretation of interface flux patterns, including those observed experimentally in [18]. Finally, the identification of a rigorous spectral refraction law has broader implications for the analysis of diffusion-wave directivity, interfacial energy-transport design, and multilayer diffusion-wave physics and engineering, extending beyond thermal diffusion to other diffusion-wave systems governed by similar parabolic operators. In addition, the rigorous modal refraction law identified here also has potential implications for physics-informed neural networks (PINNs) and spectral neural operators [21]. Because such methods increasingly rely on spectral representations of PDEs, the spectral Snell constraint may serve as a physically grounded inductive bias for studying diffusion-wave behavior in layered systems.

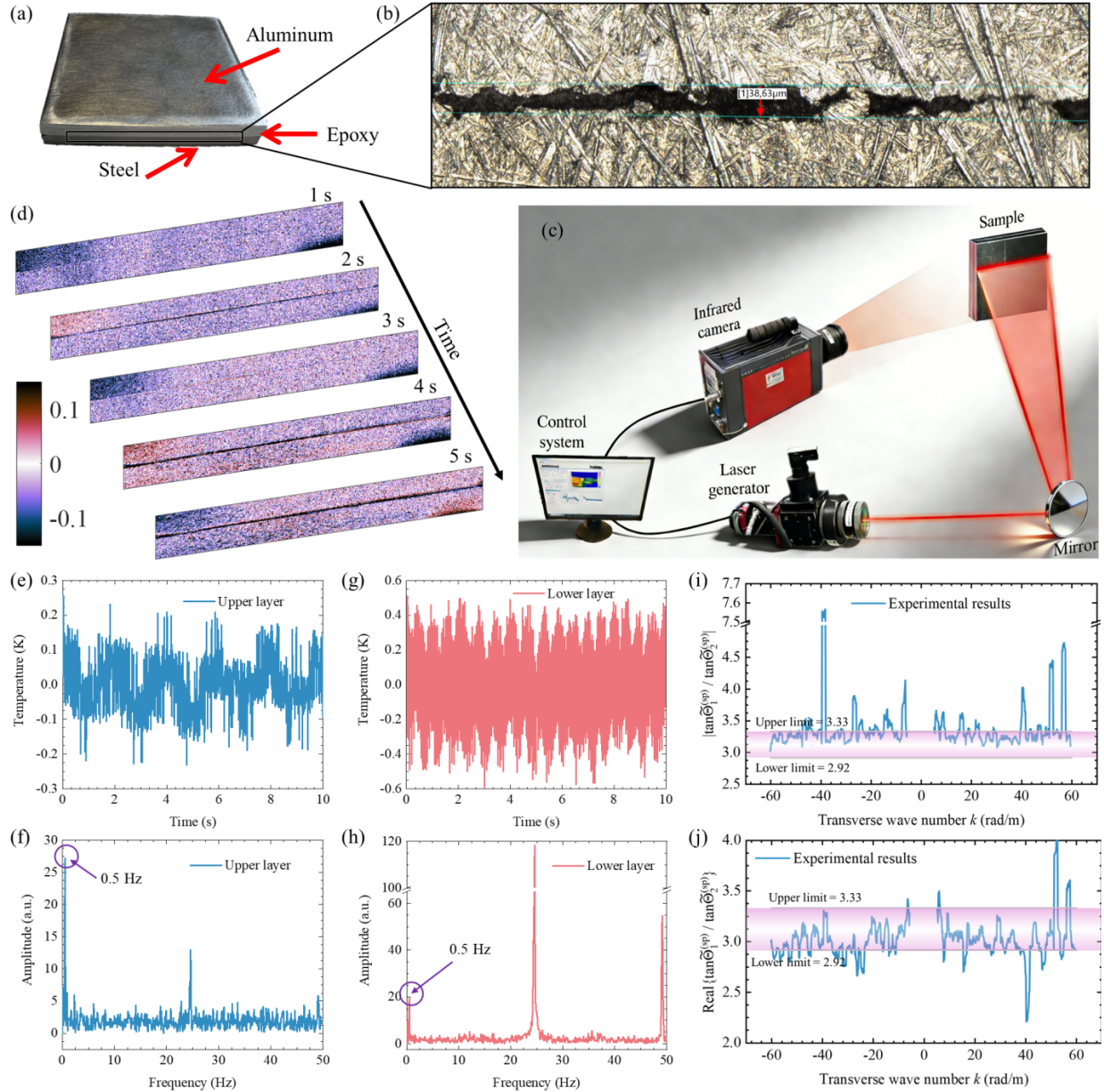

FIG. 5. Experimental validation of spectral domain Snell's law under refractive condition: (a) The tested sample (aluminum-epoxy-steel); (b) The microscopic image around the interface; (c) Experimental setup; (d) The temperature distribution varying with time; (e) The temperature variation of the upper layer; (f) The temperature in the frequency domain of the upper layer; (g) The temperature variation of the lower layer; (h) The temperature in the frequency domain of the lower layer; (i) The absolute value of $\tan\tilde{\theta}_1^{(sp)}(k,s)/\tan\tilde{\theta}_2^{(sp)}(k,s)$; (j) The real part of $\tan\tilde{\theta}_1^{(sp)}(k,s)/\tan\tilde{\theta}_2^{(sp)}(k,s)$.

*Experimental Validation*—To further validate the proposed spectral domain Snell's law, photothermal experiments were performed using two metallic samples with different thermal conductivities. A focused laser emission line was used to locally heat the front surface of two bonded metallic plates, while the resulting temperature evolution and spatial distribution were monitored on the side surface, as shown in Fig. 5(c). In practice, achieving perfect mechanical contact between two metallic plates is challenging. Residual air gaps at the interface can introduce substantial interfacial thermal resistance, severely impeding heat transport into the lower layer and perturbing intrinsic diffusion behavior. To mitigate this effect, the two metals were bonded using a thin epoxy layer and mechanically clamped for 12 hours. This procedure effectively suppresses the interfacial thermal resistance, yielding nearly continuous heat diffusion across the interface. From a physical perspective, the bonded structure constitutes an effective three-layer diffusion system: sample #1 – epoxy – sample #2. In the spectral domain, the diffusion-wave Snell's law applies successively at each interface. Specifically, for the first and second interfaces, one obtains $\frac{\tan \widetilde{\Theta}_1^{(sp)}(k,s)}{\tan \widetilde{\Theta}_2^{(sp)}(k,s)} = \frac{(\tilde{F}_{x,1}/\tilde{F}_{z,1})}{(\tilde{F}_{x,2}/\tilde{F}_{z,2})} = \frac{K_1}{K_2}$ and $\frac{\tan \widetilde{\Theta}_2^{(sp)}(k,s)}{\tan \widetilde{\Theta}_3^{(sp)}(k,s)} = \frac{(\tilde{F}_{x,2}/\tilde{F}_{z,2})}{(\tilde{F}_{x,3}/\tilde{F}_{z,3})} = \frac{K_2}{K_3}$ where $K_j$ denotes the thermal conductivity of layer $j$. By eliminating the intermediate epoxy layer, a cascade relation follows naturally, $\frac{\tan \widetilde{\Theta}_1^{(sp)}(k,s)}{\tan \widetilde{\Theta}_3^{(sp)}(k,s)} = \frac{(\tilde{F}_{x,1}/\tilde{F}_{z,1})}{(\tilde{F}_{x,3}/\tilde{F}_{z,3})} = \frac{K_1}{K_3}$, demonstrating that the spectral domain Snell's law remains valid even in the presence of a thin interfacial bonding layer. The experimental setup is shown in Fig. 5(c). The infrared camera was from Infratec ImageIR 9300 equipped with a cooled indium antimonide (InSb) focal-plane array (1280 × 1024 pixels, mid-wave infrared, 3-5 μm), providing a noise-equivalent temperature difference below 30 mK at 30 °C. The camera frame rate was set to 100 Hz, and the data were recorded for 10 s after the sample temperature approached a quasi-steady state. A diode laser (LDM 500-20, wavelength 940 ± 5 nm) was used as the excitation source, focused through a 150 mm cylindrical lens into a line spot of 17 × 0.38 mm$^2$, corresponding to a maximum power density of approximately 7.74 kW/cm$^2$. A signal generator supplied a sinusoidal modulation at 0.5 Hz, while a synchronization controller ensured that infrared acquisition was lock-in to the initial point of the modulation cycle. The samples consisted of two standard metallic plates (Fig. 5(a)), aluminum and steel, with thermal conductivity of $K_{Al} \in [140,160]$ W/(m·K) and $K_{Steel}$ = 48 W/(m·K), respectively [20]. As shown in Fig. 5(b), optical microscopy measurements performed with a Keyence VHX-S650E revealed an interfacial gap of approximately 38 μm prior to bonding, which was effectively filled with the epoxy layer. The time-resolved temperature evolution is displayed in Fig. 5(d). After removing the DC component, the temperature oscillations in the upper layer and lower layer are shown in Figs. 5(e) and (g), respectively. While the signal in the lower layer exhibits noticeable noise in the time domain, the quantity of interest is the response at the modulation frequency. Accordingly, the temporal frequency-domain signals of the upper layer and lower layer are shown in Figs. 5(f) and (h), respectively. Owing to the strong noise-rejection capability of lock-in detection, the effective harmonic component is cleanly separated from broadband noise, particularly in the lower layer (Fig. 5(h)). Of note, this processing corresponds only to the temporal frequency domain. The extracted harmonic signals were subsequently transformed into the spatial frequency domain, followed by finite-difference processing of spatial gradients to obtain the spectral flux components. The resulting spectral flux ratios are shown in Figs. 5(i) and (j), respectively. The real parts of $\tan \widetilde{\Theta}_1^{(sp)}(k,s) / \tan \widetilde{\Theta}_2^{(sp)}(k,s)$ were found to lie between 2.0 and 4.0. The mean value of the $|\tan \widetilde{\Theta}_1^{(sp)} / \tan \widetilde{\Theta}_2^{(sp)}|$ is 3.01, in close agreement with the theoretical conductivity ratio range. These results provide direct experimental confirmation that the spectral domain Snell's law holds for thermal diffusion waves, validating the theoretical framework developed in this work.

*Conclusions*—This work clarifies the physics of thermal diffusion-wave fields by identifying the representation in which wave-like interfacial relations can be rigorously defined. We show that the conclusion of Mandelis et al. [19], namely the absence of refraction and reflection, applies strictly to the composite time-domain, real-space thermal field and temporal-frequency thermal field. In contrast, by reformulating the interfacial problem in the Fourier-Laplace spectral domain, where the diffusion operator, $\partial_t - D\nabla^2$ is diagonalized with respect to the transverse wave number $k$ and the Laplace frequency $s$, we demonstrate that wave-like interfacial relations formally analogous to Snell's law naturally emerge. Each $(k,s)$ component constitutes an eigenmode of the diffusion operator with a complex propagation constant and a well-defined directional flux symmetry at material interfaces. The resulting spectral-domain Snell's relation follows directly from continuity of temperature and normal flux, and reflects a structural property of the parabolic diffusion operator. The parabolic nature of diffusion, however, prevents this modal directionality from manifesting as geometric propagation in real space. Because the observation thermal-wave response arises from the superposition of a continuum of $(k, s)$ components with distinct attenuation and dispersion rates, the inverse transformation mixes these modes, erasing any persistent refraction angle. The absence of real-space refraction therefore results not from the nonexistence of modal directionality but from irreversible spectral damping. This framework reconciles previously conflicting interpretations of thermal-wave refraction and reflection by distinguishing modal directionality in spectral space from geometric propagation in real space. Although mathematical formulations differ, the underlying physics is the same, since both descriptions are governed by the same diffusive operator. Thermal-wave modes are intrinsically lossy, processing complex wavenumbers whose imaginary parts dominate their spatial evolution. Such modes do not form an orthogonal and complete basis as in hyperbolic wave systems; the parabolic operator lacks the completeness relations that would yield Dirac-delta orthogonality (A classic example of this non-orthogonal, no-complete modal behavior is the Haynes-Shockley carrier-diffusion experiment in germanium [22], where an injected carrier packet does not propagate as a sharp wave but instead spreads into an error-function profile due to the superposition of many diffusive modes with different decay rates. This experimentally demonstrates that diffusion modes cannot maintain fixed phase or directionality, confirming that parabolic diffusion operators lack the orthogonal, complete eigenmode structure characteristic of hyperbolic wave systems). As a result, modal contributions cannot accumulate coherently to produce sustained geometric refraction, even though a well-defined spectral-domain Snell relation exists. The spectral Snell-type relation thus reflects the directional content of the modal structure, whereas the absence of observable real-space bending stems directly from the dissipative and non-complete nature of thermal-wave eigenmodes. All reported refraction-like observations, whether measured in the time-domain photothermal experiments of Bertolotti et al. [18] or the temporal-frequency diffusive-density-wave experiments of O'Leary et al. [4]), are shown to arise from the same underlying spectral-domain Snell relation. These experiments isolate narrow-band $(k, s)$ components for which modal phase and flux matching hold; once the full continuum is superposed, geometric refraction disappears, consistent with Mandelis et al [19], confirming that the spectral-domain Snell's law is the sole rigorous origin of all experimentally observed Snell-like behavior in diffusive media.

Although demonstrated here for thermal diffusion-wave fields, the spectral-domain Snell's law directly from the generic structure of linear diffusion-type equations of the form, $\frac{\partial \phi(\mathbf{r},t)}{\partial t} = D \nabla^2 \phi(\mathbf{r}, t)$, under piecewise-constant transport coefficients and interfacial flux continuity. Therefore, the same spectral interface refraction / reflection principle is expected to hold for other diffusion-wave systems.

The theoretical derivations for mass diffusion, Lindblad quantum diffusion, and electromagnetic diffusion are provided in the Supplementary document. Experimental verification of spectral refraction phenomena in these systems remains an important direction for future investigation.

P. Z. acknowledges the support from the Adolf Martens Postdoctoral Fellowship (BAM-AMF-2025-1).

**End Matter**

*Mass Diffusion*—Consider a two-dimensional ($x$-$z$) system composed of two semi-infinite layers, where layer 1 ($z > 0$) and layer 2 ($z < 0$) are characterized by diffusion coefficients $D_1$ and $D_2$, respectively. The concentration field $C(x, z, t)$ obeys Fick's law:

$$\frac{\partial C_i(x,z,t)}{\partial t} = D_i \left( \frac{\partial^2 C_i}{\partial x^2} + \frac{\partial^2 C_i}{\partial z^2} \right) + f_i(x, z, t), \quad i = 1,2 \quad (30)$$

where $f_i(x, z, t)$ denotes a volumetric source term. Applying a Laplace transform in time $t \to s$ and assuming zero initial condition, the diffusion equation becomes a Helmholtz-type equation in the Laplace domain,

$$\frac{\partial^2 \hat{C}_i}{\partial x^2} + \frac{\partial^2 \hat{C}_i}{\partial z^2} - \frac{s}{D_i} \hat{C}_i = -\hat{f}_j(x, z, s) \quad (31)$$

A spatial Fourier transform ($x \to k$) reduces Eq. (31) to a one-dimensional ODE along $z$:

$$\frac{\partial^2 \tilde{C}_i}{\partial z^2} - \kappa_i^2(k,s)\tilde{C}_i = -\tilde{f}_j(k,z,s) \quad (32)$$

where $\kappa_i(k,s) = \sqrt{\frac{s}{D_i} + k^2}$, and $\kappa_i^{-1}$ defines a complex diffusion length. one obtains spectral reflection and transmission coefficients:

$$R(k,s) = \frac{D_1\kappa_1 - D_2\kappa_2}{D_1\kappa_1 + D_2\kappa_2} \quad (33)$$
$$T(k,s) = \frac{2D_1\kappa_1}{D_1\kappa_1 + D_2\kappa_2} \quad (34)$$

According to the Fick's law, the spectral fluxes are defined as:

$$\tilde{J}_{x,j} = -D_j(ik)\tilde{C}_j \quad (35)$$
$$\tilde{J}_{z,j} = -D_j \partial_z \tilde{C}_j \quad (36)$$

The refraction law in the spectral domain reads:

$$\frac{\tan \Theta_1^{(\text{sp})}(k,s)}{\tan \Theta_2^{(\text{sp})}(k,s)} = \frac{D_1}{D_2} \quad (37)$$

*Quantum Diffusion Waves in Lindblad Systems*—Consider a quantum system described by a density matrix $\rho$ obeying the Lindblad master equation,

$$\frac{\partial \rho}{\partial t} = -\frac{i}{\hbar}[H,\rho] + \sum_{n,m} h_{nm}(A_n \rho A_m^\dagger - \frac{1}{2}\{A_m^\dagger A_n, \rho\}) \quad (38)$$

where $H$ is the Hamiltonian, $\hbar$ is the reduced Plank constant, $[H,\rho] = H\rho - \rho H$ represents the coherent evolution of the system according to Schrödinger dynamics. The matrix $h_{nm}$ is the Hermitian positive-definite Kossakowski matrix, and $A_n$ and $A_m$ are Lindblad (jump) operators, with $\{X,Y\} = XY + YX$ denoting the anticommutator. The dissipator is defined as $\mathcal{D}[\rho_j] = \sum_{n,m} h_{nm}(A_n \rho A_m^\dagger - \frac{1}{2}\{A_m^\dagger A_n, \rho\})$. For regimes dominated by dissipation ($\|\mathcal{D}[\rho_j]\| \gg \|[H,\rho]/\hbar\|$), the Hamiltonian term can be neglected, and the master equation reduces to $\frac{\partial \rho_j}{\partial t} \simeq \mathcal{D}[\rho_j]$. For spatially-localized jump operators, $A_n \sim |r_n\rangle\langle r_n|$, consider the diagonal elements (population densities) of the density matrix: $\rho_j(x,z,t) \equiv \langle r|\rho_j|r\rangle$. Under small displacement, continuous-space approximation, the dissipator can be approximated by a diffusion operator, $\mathcal{D}[\rho_j] \simeq D_j \nabla^2 \rho_j(x,z,t)$, where the effective Lindblad decoherence/diffusion coefficient in layer $j$ is $D_j \simeq \frac{1}{2}\sum_{n,m} h_{nm}^{(j)} |r_n^{(j)} - r_m^{(j)}|^2$, with $r_n$ and $r_m$ denoting the spatial positions associated with the jump operators $A_n$ and $A_m$, respectively. For localized excitations, e.g., photon injection, we include a source $f_j(x,z,t) = \delta(x-x_0)\delta(z-z_0)\delta(t)$, leading to the layered Lindblad diffusion equation:

$$\partial_t \rho_j(x,z,t) = D_j \nabla^2 \rho_j(x,z,t) + f_j(x,z,t), \quad j=1,2 \quad (39)$$

Applying Fourier transform in $x$ ($x \to k$) and Laplace transform in $t$ ($t \to s$) gives

$$D_j(\partial_{zz}\tilde{\rho}_j - \kappa_j^2 \tilde{\rho}_j) = -\tilde{f}_j(k,z,s) \quad (40)$$

with the complex quantum diffusion wavenumber: $\kappa_j^2(k,s) = k^2 + \frac{s}{D_j}$. The spectral reflection and transmission coefficients read:

$$R(k,s) = \frac{D_1\kappa_1 - D_2\kappa_2}{D_1\kappa_1 + D_2\kappa_2} \quad (41)$$
$$T(k,s) = \frac{2D_1\kappa_1}{D_1\kappa_1 + D_2\kappa_2} \quad (42)$$

yielding the spectral-domain Snell's law for Lindblad diffusion waves:

$$\frac{\tan \tilde{\Theta}_1^{(\text{sp})}(k,s)}{\tan \tilde{\Theta}_2^{(\text{sp})}(k,s)} = \frac{D_1}{D_2} = \frac{\sum_{n,m} h_{nm}^{(1)} |r_n^{(1)} - r_m^{(1)}|^2}{\sum_{n,m} h_{nm}^{(2)} |r_n^{(2)} - r_m^{(2)}|^2} \quad (43)$$

*Maxwell Electromagnetic Diffusion Waves Systems*—Consider two semi-infinite conducting layers in the $x$-$z$ plane. The upper layer ($0 \leq z \leq \infty$) has conductivity $\sigma_1$ and permeability $\mu_1$, while the lower layer ($z < 0$) has $\sigma_2$ and $\mu_2$. In the high-conductivity (diffusion-wave) limit, where displacement currents are negligible compared to conduction currents ($\sigma \gg \omega\varepsilon$, $\varepsilon$ is the dielectric constant) the electric field satisfies a diffusion-type equation:

$$\frac{\partial E_j}{\partial t} = \frac{1}{\mu_j \sigma_j}\left(\frac{\partial^2 E_j}{\partial x^2} + \frac{\partial^2 E_j}{\partial z^2}\right) + f_j(x,z,t), \quad j=1,2 \quad (44)$$

where $f_j = \delta(x-x_0)\delta(z-z_0)\delta(t)$ represents a localized current source. Applying a Fourier transformation in $x$ ($x \to k$) and a Laplace transformation in $t$ ($t \to s$) reduces Eq. (44) to a 1D Helmholtz-type ODE along $z$

$$\frac{1}{\mu_j \sigma_j}\left(\partial_{zz} - \kappa_j^2(k,s)\right)\tilde{E}_j(k,z,s) = -\tilde{f}_j(k,z,s) \quad (45)$$

where the complex diffusion wave number is $\kappa_j^2(k,s) = k^2 + s\mu_j\sigma_j$. The spectral reflection and transmission coefficients read:

$$R(k,s) = \frac{\sigma_2\mu_2\kappa_1 - \sigma_1\mu_1\kappa_2}{\sigma_2\mu_2\kappa_1 + \sigma_1\mu_1\kappa_2} \quad (46)$$
$$T(k,s) = \frac{2\sigma_2\mu_2\kappa_1}{\sigma_2\mu_2\kappa_1 + \sigma_1\mu_1\kappa_2} \quad (47)$$

Defining the spectral energy flux analogous to classical diffusion:

$$\tilde{J}_{x,j} = -\frac{1}{\mu_1 \sigma_1}(ik)\tilde{E}_j \quad (48)$$

$$\tilde{J}_{z,j} = -\frac{1}{\mu_2 \sigma_2}\partial_z \tilde{E}_j \quad (49)$$

yielding the spectral-domain Snell's law for electromagnetic diffusion waves:

$$\frac{\tan \tilde{\Theta}_1^{(sp)}(k,s)}{\tan \tilde{\Theta}_2^{(sp)}(k,s)} = \frac{\sigma_2 \mu_2}{\sigma_1 \mu_1} \quad (50)$$